\documentclass[12pt]{article}
\usepackage{latexsym}

 \csname
@addtoreset\endcsname{equation}{section}

\newcommand{\ghe}[3]{$\stackrel{\textstyle #1}{\scriptstyle (#2,#3)}$}
\newcommand{\lre}{\multicolumn{3}{c}{$\stackrel{*}{\longleftrightarrow}$}}

\begin{document}
\begin{titlepage}
\begin{flushright}
NYU-TH/00/09/11 \\
SNS-PH/00-15 \\
hep-th/0010052
\end{flushright}
\vfill
\begin{center}
{\LARGE\bf Notes on the Quantization \\ of the Complex Linear
  Superfield}    \\ 
\vskip 27.mm  \large
{\bf  P.~A.~Grassi~$^a$, G.~Policastro~$^{a,b}$,  and  M.~Porrati~$^a$}\\
\vfill
{\small \it
$(a)$ Physics Department, New York University, \\ 
4 Washington Place, New York,  NY 10003,  USA \\ \vspace{.2cm}
$(b)$ Scuola Normale Superiore, \\ 
Piazza dei Cavalieri  7, 56100, Pisa, Italy \\}
\end{center}
\vfill

\begin{center}
{\bf ABSTRACT}
\end{center}
\begin{quote}
  The quantization of the complex linear superfield requires an
  infinite tower of ghosts.  Using the Batalin-Vilkovisky technique,
  Grisaru, Van Proeyen, and Zanon have been able to define a correct
  procedure to construct a gauge-fixed action.  We generalize their
  technique by introducing the Lagrange multipliers into the
  non-minimal sector and we study the BRST cohomology.
  We show how the physical subspace is singled out. Finally, we
  quantize the model in the presence of a background and of a quantum
  gauge superfield.  \vfill \hrule width 5.cm \vskip 2.mm
\end{quote}
\begin{flushleft}
October 2000
\end{flushleft}
\end{titlepage}


\section{Introduction}
\label{introduction}

The correct quantization of systems with an infinitely reducible gauge
symmetry is a longstanding problem. Recently, a new technique for such
models has been developed.  The authors of \cite{GPZ} showed that a
correct implementation of the BV-BRST \cite{BV,BRST}
formalism provides a meaningful quantization procedure for the Complex
Linear Superfield.

The complex linear superfield is a constrained field and provides a
dual description of the chiral superfield with a different set of
auxiliary fields ~\cite{gates,superspace,WZNW,YMSG,Deo}.  Solving the
constraints in terms of an unconstrained spinor superfield, the
redundant unphysical degrees of freedom are removed by new gauge
symmetries. The latter, however, turn out to be reducible: there exist
non-trivial zero modes of the gauge transformations.  This implies
that ghosts for ghosts should be introduced in order to have the
correct number of degrees of freedom. Furthermore, unlike the case of
the antisymmetric tensor field~\cite{AT}, the transformations are
infinitely reducible. This means that an infinite tower of ghost
fields is indeed required in order to quantize the model covariantly.

In ref.~\cite{GPZ}, it was shown that, by a simple diagonalization
technique, the propagator of the fields can be constructed
algorithmically provided a well defined way of fixing the gauge for
the tower of ghosts is given. Essentially, the technique is based on a
sequence of canonical transformations designed to disentangle the
different generations of ghost fields. The sequence of transformation
is constructed in such a way that the gauge fixing of the $n^{\rm th}$ order
can be done without affecting the $(n-1)^{\rm th}$ order.  However, those
diagonalizations, performed by non-local redefinitions, might cause
severe problems. We will show that the theory is local and the
BRST cohomology changes drastically as a consequence of those
transformations.

The case of the linear superfield discussed in refs. \cite{GPZ,PZ} is
fairly simple. The ungauged version of that model possesses a tower of
free ghost fields which are zero modes of the {\it off-shell} gauge
transformations.  Therefore, as shown in \cite{GPZ}, the BV technique
can be used for a correct quantization procedure. Since the ghost
fields are free, one can easily compute the functional integral and
the correlation functions. The gauged version, on the other hand, is a
model with a reducible gauge symmetry where the ghost fields interact
with Yang-Mills gauge superfield through supercovariant derivatives.
Some years ago, P.  Townsend proposed a way to handle such a situation
\cite{3lectures}: he suggested to introduce redefined ghost fields
such that, in terms of these new variables, the ghosts for ghosts do
not interact with the gauge fields. Therefore, by choosing a suitable
gauge fixing, the procedure proposed in \cite{GPZ} can be used again
to quantize the model. This is one of the aims of the present paper.
We also extend Townsend's idea to the case of quantum gauge fields
coupled to superfields.

Our interest in the quantization of the linear superfield is mainly
due to the extension of the technique developed in \cite{GPZ,PZ} 
to other models with an infinitely reducible gauge symmetry, as 
the Casalbuoni-Brink-Schwarz superparticle
\cite{Brink-Schwarz,Siegel:1983hh,Lindstrom_BS,GH_BS,superparticle},
the Green-Schwarz superstring
\cite{green3,hiddengh,GSST,bergshoeff,GH,Lindstrom}, and string
field theory \cite{SFT}. 

The cases of the Casalbuoni-Brink-Schwarz superparticle and of the
Green-Schwarz superstring have some common features with the gauged
linear superfield model. In particular, in all these cases the ghosts
of the $\kappa$-symmetry \cite{Siegel:1983hh} are interacting with the
{\it physical} fields.  Furthermore, the $\kappa$-transformations are
reducible only {\it on-shell}.  This requires a slightly different
quantization technique which has been discussed in \cite{porr_2}. It
has been shown there that, by a suitable canonical transformation, the
tower of ghost for the $\kappa$-symmetry coincides with an off-shell
tower of ghosts which describe the same degrees of freedom in suitable
variables.  Furthermore, the Lorentz gauge-fixing type proposed in
\cite{GPZ,PZ} cannot be used for the superparticle or the superstring.
Instead, the Landau gauge fixing proposed in \cite{Lindstrom_BS} can
be implemented by using an extended non-minimal sector.

In the present paper, we provide a small generalization of the
technique of \cite{GPZ} which takes into account the Lagrange
multipliers. We also identify the correct number of degrees of freedom
and, consequently, the correct physical subspace with the BRST
cohomology. We
compute the cohomology for the gauge unfixed theory and for the gauge
fixed one. In addition, we show that the naive application of the
technique of \cite{GPZ} does not provide the correct physical
subspace, which can be recovered nevertheless by a careful handling of
the non-local canonical transformations.

Here, we provide a complete discussion of the realization of
Townsend's idea.  In section (\ref{review}) we review the classical
theory and we give a brief account on the BV formalism. For
convenience, we follow the notation and the conventions of \cite{GPZ}.
In section (\ref{coho}) we show that, by introducing suitable
variables for the ghost fields, the technique of \cite{GPZ} can be
used to quantize the model. Furthermore, we discuss the BRST cohomology, 
and we show that the quantized
theory describes the correct degrees of freedom. In particular, we
adopt a simpler model (Section (\ref{toy_model})) to show that the
procedure outlined in the previous section can be implemented to all
levels of ghosts. In section (\ref{gauged}) we discuss the gauged
version of the model. Section (\ref{conclusion}) summarizes our
conclusions. The Appendices contain additional technical developments.

\newpage

\section{The classical theory and  BV formalism}
\label{review}

\subsection{The classical action and its gauge symmetry}
\label{classical}

The kinetic action for a complex linear superfield $\Sigma$, $\bar\Sigma$,
with $\bar D^2 \Sigma = D^2 \bar\Sigma =0$ is 
\begin{equation}
  S=- \int d^4x d^4 \theta ~ \bar\Sigma \Sigma~.
\end{equation}
The equivalence of the descriptions of the scalar
multiplet by the linear superfield $\Sigma$ and by the chiral superfield
$\Phi$ can be exhibited by means of a duality transformation, starting
with the action
\begin{equation}
  S_D = -\int d^4x d^4 \theta ~ [ \bar{\Sigma} \Sigma +\Phi \Sigma
+\bar\Phi \bar{\Sigma} ]
\end{equation}
(with { \em unconstrained} $\Sigma$ and chiral
$\Phi$) \cite{superspace}.  Using the equations of motion to eliminate
the superfields $\Sigma, \bar\Sigma$, leads to the usual chiral superfield
action. Eliminating instead the superfields $\Phi$, $\bar\Phi$ (whose
equations of motion impose the linearity constraint $\bar D^2 \Sigma = D^2
\bar\Sigma =0 $) leads to the linear superfield action.

The linearity constraint can be solved in terms of an unconstrained
spinor superfield and its complex conjugate by
\begin{equation}
\Sigma = \bar D _{\dot\alpha} \bar\sigma^{\dot\alpha} ~~~~,~~~~
\bar\Sigma = D_\alpha \sigma^\alpha \ , 
\end{equation}
and the action becomes
\begin{equation}
S_{cl}= -\int d^4x d^4 \theta ~~D_\alpha \sigma^\alpha \cdot \bar
D_{\dot\alpha} \bar\sigma^{\dot\alpha}
=-\int d^4x d^4 \theta  ~\sigma^\alpha D_\alpha \bar D_{\dot\alpha}
\bar\sigma^{\dot\alpha} \, .
\label{Scl}
\end{equation}
The solution of the constraint has introduced some gauge invariance,
since the general spinor superfield $\sigma^\alpha$ has more
components than $\Sigma$. Thus 
the operator $D_\alpha \bar D_{\dot\alpha}$ is not invertible.
We view $\Sigma$ as the field strength of the gauge field $\sigma^\alpha$.

The above action is invariant under the variation $\delta
\sigma^\alpha = D_\beta \sigma^{(\alpha \beta)}$ with unconstrained
{\em symmetric} (as indicated by the brackets) bispinor gauge
parameter, since $D_\alpha D_\beta$ is antisymmetric in the indices.
However, this variation has zero modes, $\delta \sigma^{(\alpha
  \beta)} = D_\gamma \sigma^{(\alpha \beta \gamma)}$ with the new
parameter symmetric in its indices.  Similarly, we find zero modes
$\delta \sigma^{(\alpha\beta \gamma)} = D_\delta \sigma^{(\alpha\beta
  \gamma \delta)}$, and so on. Proceeding in this manner one discovers
an infinite chain of transformations with zero modes which, upon
quantization, leads to an infinite tower of ghosts. This comes about
because at every step the ghosts have more components than are
necessary to remove gauge degrees of freedom, an apparently
unavoidable situation if one wants to maintain manifest Lorentz
invariance.  It is this feature which makes the quantization of the
complex linear superfield difficult.


\subsection{Symmetries and the anti-bracket formalism}
\label{BV}

The essential ingredients in the Batalin-Vilkovisky (BV)
\cite{BV,cano,gomis,henne_libro,anti_coho} formalism are the anti-fields and
anti-brackets. For any field $\phi ^A$, one introduces an anti-field
$\phi ^*_A$.  They have statistics opposite to their corresponding
fields and a ghost number, assigned such that for the classical fields
$gh(\Phi) =0$, the (extended) action has ghost number zero, and for
all fields $gh(\phi^*_A )= -gh(\phi ^A)-1$.

The {anti-bracket} 
of two functions $F$ and $G$ is defined by
\begin{equation}\label{abracket}
\left( F, G \right) = 
\frac{\partial_r }{\partial\phi ^A}  F \frac{\partial_l }{\partial\phi
  ^*_A} G -  
\frac{\partial_r }{\partial\phi ^*_A} F  \frac{\partial_l
  }{\partial\phi ^A} G \, ,   
\end{equation}
where $\partial_{r/l}$ denote the derivative from the left or from the
right.  They satisfy graded commutation, distribution and Jacobi
relations \cite{BV}.  For these brackets, fields and anti-fields behave
as coordinates and their conjugate momenta
\begin{equation}\label{eq:brackffs}
(\phi^A,\phi^B)=0\ ;\hspace{0.5cm}
(\phi^*_A,\phi^*_B)=0;\hspace{0.5cm}
(\phi^A,\phi^*_B)=\delta^A_B.
\end{equation}
The extended action $S$ is the solution of the master equation 
\begin{equation}\label{master}
(S,S)=0\,,
\end{equation}
with the boundary condition that the classical action $S_{\rm class}$
coincides with minimal action when the ghost fields are set to zero
and the anti-fields $\phi^*_A$ are coupled to the gauge
transformations of the classical fields.  Notice that there is a
natural grading among the fields and the anti-fields, namely the
antighost number \cite{gomis}. This allows a convenient decomposition
of the extended action and, accordingly, the master equation can be
easily solved.

Canonical transformations are an important part of the formalism
\cite{cano}.  They preserve the anti-bracket structure
(\ref{abracket}): calculating the anti-brackets in the old or new
variables is the same, or, in other words, the new variables also
satisfy (\ref{eq:brackffs}). Therefore also the master
equation $(S,S)=0$ is preserved under such transformations.

Canonical transformations from $\{\phi\, \phi ^*\}$ to $\{\phi'\,
\phi'^*\}$, for which the matrix $\left. \frac{\partial_r
    \phi^B}{\partial \phi'^A}\right|_{\phi'^*}$ is invertible, can be
obtained from a fermionic generating function $F(\phi,\phi'^*)$. The
transformations are defined by
\begin{equation} \phi'^A=\frac{\partial F(\phi , \phi'^*)
}{\partial \phi'^*_A}\hspace{2cm}
\phi^*_A=\frac{\partial F(\phi , \phi'^*)}{\partial \phi ^A}.
\label{Fcan}\end{equation}

\begin{itemize}
\item Point transformations are the easiest ones. These are just
  redefinitions of the fields $\phi '^A=f^A(\phi )$. Their generating
  function is $F=\phi '^*_A f^A(\phi)$ which thus determines the
  corresponding transformations of the anti-fields. The latter replace
  the calculations of the variations of the new variables.
\item Adding the BRST transformation of a function, $s\Psi(\phi )$, to
  the action is obtained by a canonical transformation with $F=\phi
  '^*_A\phi ^A+ \Psi (\phi )$. The latter gives
\begin{equation}
\phi'^A = \phi^A \ ;\qquad
\phi ^*_A =\phi'^*_A + \partial _A \Psi (\phi ).
\label{gfermion}\end{equation}
and by means of these transformations it is possible to implement the
gauge fixing conditions as a canonical change of variables.
\end{itemize}
The canonical transformations leave by definition the master equation
invariant, and because they are non-singular, they also keep the
properness requirement on the extended action. We will discuss this
point at length later on. Of course in the new variables, we do not
see the classical limit anymore. But the most important property is
that the BRST cohomology \cite{anti_coho} is not changed.

This minimal solution, or the so-called extended action, coincides
with the classical action  when the anti-fields are set to zero. The
anti-field dependent terms encode the structure of the symmetry algebra:
the terms linear in the anti-fields generate the quantum BRST
transformation of the fields, according to the formula
\begin{equation}\label{omega}
  \gamma \Phi^A = (\Phi^A, S)|_{\phi^*=0}
\end{equation}
while the higher-order terms contain the non-closure functions of the
algebra. $\gamma$ defines a coboundary operator and the space of physical 
degrees of freedom is identified with the cohomology group $H^*(\gamma)$ 
with zero ghost number. This defines the so-called weak BRST cohomology \cite{henne_libro}, 
that is modulo the equations of motion.  

Alternatively and equivalently, one can compute the BRST cohomology 
singled out by means of the differential 
\begin{equation}\label{esse}
  s \Phi^A = (\Phi^A, S)\,,~~~~  s \Phi^*_A = (\Phi^*_A, S)\,. 
\end{equation}
In the case of the BRST cohomology, the physical subspace is
selected by decomposing the differential $s = \gamma + \delta$,
precisely into the BRST differential $\gamma$ and the Koszul-Tate
resolution $\delta$, and identifying the physical degrees of freedom
by the cohomology $H^*(\gamma,H^*(\delta))$. As shown in
\cite{anti_coho}, the two cohomological groups coincide.



\subsection{GPZ technique for infinite ghost towers}
\label{GPZ}

Following the conventions of \cite{GPZ}, the minimal classical action
(\ref{Scl}) containing all the minimal zero modes and the couplings
with their corresponding antifields is written as
\begin{equation}
S_{min}= \bar
\sigma^{\dot \alpha}\bar D_{\dot \alpha} D_\alpha\sigma^\alpha\  + 
\sum^{\infty}_{i=1}\sigma_{A_i}^* D_\beta \sigma^{(\beta A_i)} +  
\sum^{\infty}_{i= 1} \bar\sigma^{(\dot \beta \dot A_i)}  \bar D_{\dot
  \beta} \bar\sigma_{\dot A_i}^* 
\, , \label{min}
\end{equation}
where $A_i$ denotes the symmetrized set of indices $(\alpha_1,
\dots,\alpha_2)$.  

According to the BV formalism, and indicating the antifields of the
antighosts respectively by $b^{*\alpha}_{\dot \alpha}$ and $\bar
b^{*\dot \alpha}_\alpha$, we add to $S_{min}$ the non--minimal term
\begin{equation}\label{Snm1}
S_{nm,1}= \bar b^{*\dot
\alpha}_\alpha b^{*\alpha}_{\dot \alpha}\,,  
\end{equation}
and we perform the canonical transformations generated by the gauge fermion
\begin{equation}\label{Psi1}
\Psi_1= b_\alpha^{\dot \alpha} \bar D_{\dot
\alpha}\sigma^\alpha +\bar
\sigma^{\dot \alpha}D_\alpha\bar b_{\dot \alpha}^\alpha \, , 
\end{equation}
which shifts $b^*,\sigma^*$ antifields of the antighost and of
$\sigma$ field, respectively (and the corresponding hermitian
counterparts). Thus, the $0^{\rm th}$-level quantized action reads
\begin{eqnarray}
S_1&=& S_{Q,1}+S_{0,1}+S_{*,1}+ 
\sum^{\infty}_{i=1}\sigma_{A_i}^* D_\beta \sigma^{(\beta A_i)} +  
\sum^{\infty}_{i= 1} \bar\sigma^{(\dot \beta \dot A_i)}  \bar D_{\dot
  \beta} \bar\sigma_{\dot A_i}^* 
\,, \nonumber\\
S_{Q,1}&=& \bar \sigma^{\dot
\alpha}i\partial_{\dot \alpha\alpha} \sigma^\alpha \,,\nonumber\\
S_{0,1}&=&
 b_\alpha^{\dot \alpha}\bar D_{\dot \alpha} D_\beta
\sigma^{(\beta\alpha)} + {\rm h.c.}\,,
\nonumber\\
S_{*,1}&=&  \bar b^{*\dot \alpha}_\alpha b^{*\alpha}_{\dot \alpha}+
\bar b^{*\dot \alpha}_\alpha \bar D_{\dot \alpha}  \sigma^\alpha +
 \bar \sigma^{\dot \alpha} D_\alpha b^{*\alpha}_{\dot \alpha} \, .
\label{Slevel1}
\end{eqnarray}
As explained in \cite{GPZ}, in order to remove the couplings between
the antifields $b^*, \bar b^*$ and the $0^{\rm th}$-level quantized
fields, it is convenient to redefine the variables by a canonical
transformations generated by the fermion
\begin{equation}
F(\Phi , \tilde\Phi^*) =\Phi^A \tilde\Phi^*_A -
\tilde{\bar b}{}^{*\dot \alpha}_\alpha \bar D_{\dot \alpha}\frac{1}{\Box}
i\partial^{\alpha\dot \beta}\tilde{\bar \sigma}^*_{\dot \beta}-
\tilde{\sigma}^*_{ \beta}\frac{1}{\Box}
i\partial^{\beta\dot\alpha} D_ \alpha\tilde{ b}^{* \alpha}_{\dot\alpha}\ .
\label{Fdiagsigma}
\end{equation}
We notice that those transformations are non-local and they affect
both the couplings between $b^*, \bar b^*$ with $\sigma$ and $\bar
\sigma$ and the minimal terms involving $\sigma^*, \bar \sigma^*$.
These redefinitions might cause severe problems for the locality of
the gauge fixed theory. In the forthcoming sections, we analyze
carefully the problem and we show how the non-locality is essential to
compute the BRST cohomology.

Finally, in order to fix the gauge for ghosts and for extra ghosts, 
one has to add non-minimal terms to the action
\begin{eqnarray}
S_{0}&=&S_{cl}+ \sum^{\infty}_{i=1}\sigma_{A_i}^* D_\beta
\sigma^{(\beta A_i)} +   
\sum^{\infty}_{i= 1} \bar\sigma^{(\dot \beta \dot A_i)}  \bar D_{\dot
  \beta} \bar\sigma_{\dot A_i}^*  
+ S_{f,L}+S_{f,R}\nonumber\\
S_{f,L}&=&\sigma^*_{\alpha_1\alpha_2}C^{\alpha_1\alpha_2}\lambda+
\sigma^*_{A_2\alpha_3} C^{\alpha_2\alpha_3}\lambda^{\alpha_1}\nonumber\\
&& +\sigma^*_{A_3\alpha_4} C^{\alpha_3\alpha_4}\lambda^{A_2}
+\varsigma^* C_{\alpha_1\alpha_2}\lambda^{\alpha_1\alpha_2}
+ \ldots \ ,  \label{Sminf}
\end{eqnarray}
where $S_{f,R}$ is the complex conjugate of $S_{f,L}$. We do not
repeat here the considerations of \cite{GPZ} which justified the
introduction of those new ghosts and their corresponding antighosts.
However, we would like to underline that this is a specific feature of
the present model. In the case of the Casalbuoni-Brink-Schwarz
superparticle and of the Green-Schwarz superstring, this phenomenon
does not occur, since the Lorentz representation of $\kappa$-symmetry
ghosts does not change from one level to another.

The gauge fixing of ghosts and of other non-minimal fields is
obtained in the same way as the $0^{\rm th}$-level, i.e. by introducing the
non-minimal couplings
\begin{eqnarray}\label{S3star}
S_{nm,2}&=&
d^{*\dot \alpha}_{A_2}b^{*A_2}_{\dot \alpha}+
\nu^*\mu^* + h.c.\nonumber\\
S_{nm,3}&=& \bar e^{*\dot A_2}_{A_2}
e_{\dot A_2 }^{*A_2} +
\left( d^{*\dot \alpha}_{A_3}b^{*A_3}_{\dot \alpha}+
  \nu^*_\alpha \mu^{*\alpha}+d^*b^*+ {\rm h.c.}\right) \nonumber\\
&&-\bar \rho^*_\alpha i\partial^{\alpha\dot \alpha} \rho^*_{\dot \alpha}
-\bar \rho'_\alpha i\partial^{\alpha\dot \alpha}\rho'_{\dot \alpha} \, , 
\end{eqnarray}
and by performing the following canonical transformations 
\begin{eqnarray}
  \Psi_2&=&b_\alpha^{\dot \alpha}
  D_\beta d^{(\beta\alpha)}_{\dot \alpha}+b_\alpha^{\dot \alpha}
  i\partial^\alpha{}_{\dot
    \alpha}\nu
  +b_{A_2}^{\dot \alpha}\bar D_{\dot
    \alpha}\sigma^{A_2}
  +{1 \over 2}\mu C_{\alpha\beta}\sigma^{\beta\alpha} +{\rm h.c.} 
  \nonumber \\
  \Psi_3&=&b^{\dot \alpha}_{A_3}\bar D_{\dot \alpha}
  \sigma^{A_3}+{2 \over 3}\mu_{\alpha_1} C_{\alpha_2\alpha_3}
  \sigma^{A_2\alpha_3}+b^{\dot \alpha}_{A_2}\left(
    D_{\alpha_3} d_{\dot \alpha}^{A_3} +i\partial_{\dot \alpha}
    ^{\alpha_1} \nu^{\alpha_2}+ i\partial^{\alpha_1}_{\dot \alpha}
    D^{\alpha_2} d\right) \nonumber\\
  && +e_{A_2}^{\dot A_2}\bar D_{\dot \alpha_1}
  d_{\dot \alpha_2}^{A_2}
  -2 \rho^{\dot \alpha}\bar D_{\dot \alpha}\nu   + {\rm h.c.}
  \ . 
\end{eqnarray}
Again, to remove the mixing between the fields of the $1^{\rm st}$-level 
and those of the $2^{\rm nd}$-level, one has to perform
suitable diagonalizations, which are similar to the non-local
transformations generated by (\ref{Fdiagsigma}). The non-minimal terms
of (\ref{S3star}) contain also the Nielsen-Kallosh ghosts \cite{NKgh}, 
necessary in order to compensate the extra-propagating modes as discussed in
\cite{GPZ}.

In the same spirit as above, one continues to fix all the ghost kinetic
terms.  Since we are not interested to re-derive the full result of
\cite{GPZ}, we will refer to that paper also for the complete
resulting action.  In the forthcoming sections, we will show how the
same result can be obtained by introducing suitable Lagrange
multipliers associated with the antighost fields and we construct the
corresponding canonical transformations to decouple different levels
before the gauge-fixings.


\section{BRST cohomology and physical degrees of freedom}
\label{coho}

\subsection{First levels}
\label{first}

In order to provide a clear description of the cohomology computation
for the gauge unfixed and gauge fixed theory, we introduce the
Lagrange multipliers associated to the antighost fields.  The counting
of classical degrees of freedom does not change, but the computation
of the gauge-fixed degrees of freedom is simplified. Furthermore, as
explained in the introduction, for some models a suitable gauge fixing
can be only implemented by means of the Lagrange multipliers (cf.
\cite{porr_2}).

We introduce the Lagrange multipliers $\beta^\alpha_{\dot\alpha},
\delta^{A_2}_{\dot\alpha}, \beta_{A_2}^{\dot\alpha}, \dots$
associated with the antighost fields and with the extra ghost fields
$b^\alpha_{\dot\alpha}, d^{A_2}_{\dot\alpha}, b_{
  A_2}^{\dot\alpha}, \dots $ displayed in table (\ref{tab1}).
\footnote{Following \cite{GPZ} the symbol $\sigma^{\alpha\beta}$
  contains both the symmetric part (the minimal ghost) and the
  antisymmetric part (the non-minimal ghost): $\sigma^{\alpha \beta}=
  \sigma^{(\alpha \beta)} + C^{[\alpha \beta]}\lambda $.}
The Lagrange multipliers obey: 
\begin{eqnarray}
  \label{lag_mul}
&&  
s \, b^\alpha_{\dot\alpha} = \beta^\alpha_{\dot\alpha}\,,~~~
s \, d^{A_2}_{\dot\alpha} = \delta^{A_2}_{\dot\alpha}\,, ~~~
s \, \nu = n\,, ~~~
s \,  b_{\dot A_2}^{\dot\alpha} = \beta_{\dot A_2}^{\dot\alpha}\,,~~~ 
s \, \mu = m\,, ~~~
\dots\,, 
 \nonumber  \\
&&
s \, \beta^\alpha_{\dot\alpha} = 0\,,~~~~
s \, \delta^{A_2}_{\dot\alpha} = 0\,, ~~~~~~
s \, n = 0, ~~~~
s \, \beta_{\dot A_2}^{\dot\alpha} = 0\,,~~~~~
s \, m = 0\,, ~~~
\dots\,, 
\end{eqnarray}
where $s$ is the generators of the classical BRST transformations. As
a convention, we denote with a Greek letter the Lagrange multiplier if
the corresponding ghost fields is denoted by a Latin letter, and 
vice-versa.  In table (\ref{tab2}) the Lagrange multipliers are shown,
together with their spinorial indices and their quantum numbers.

Using the Lagrange multiplier, in place of (\ref{Snm1}), we add the
non-minimal terms
\begin{equation}
S_{nm,1}= s \left(  \bar b^{*\dot \alpha}_\alpha b^{\alpha}_{\dot \alpha} + 
\bar b^{\dot \alpha}_\alpha    b^{*\alpha}_{\dot \alpha}
\right) = \bar b^{*\dot \alpha}_\alpha \beta^{\alpha}_{\dot \alpha}  +  
\bar \beta^{\dot \alpha}_\alpha    b^{*\alpha}_{\dot \alpha}\,,  \label{nm1}
\end{equation}
and we generate the canonical transformations on the fields of the
minimal action $S_{min}$ of eq.~(\ref{Sminf}) by means of the gauge
fermion
\begin{equation}
\Psi_1= 
b_\alpha^{\dot \alpha} 
\left( \bar D_{\dot \alpha}\sigma^\alpha + k \beta^{\alpha}_{\dot
    \alpha} \right)  + 
\left( \bar \sigma^{\dot \alpha} 
 D_\alpha  + \bar k \bar \beta^{\dot
    \alpha}_\alpha \right) \bar b_{\dot \alpha}^\alpha \,. 
\label{Psi1_new}
\end{equation}
Here, $k$ is a complex gauge parameter (cf.~\cite{GPZ} for a
discussion on gauge parameters in the quantization of the complex
linear superfield). The resulting action is given by
\begin{eqnarray}
S'_1&=& S'_{Q,1}+S_{0,1}+S'_{*,1}+ 
\sum^{\infty}_{i=1}\sigma_{A_i}^* D_\beta \sigma^{(\beta A_i)} +  
\sum^{\infty}_{i= 1} \bar\sigma^{(\dot \beta \dot A_i)}  \bar D_{\dot
  \beta} \bar\sigma_{\dot A_i}^* 
\,, \nonumber\\
S'_{Q,1}&=&  
\bar\sigma^{\dot \alpha}\bar D_{\dot \alpha} D_\alpha\sigma^\alpha\ +
\bar \beta^{\dot \alpha}_\alpha \bar D_{\dot \alpha}\sigma^\alpha + 
\bar \sigma^{\dot \alpha} D_\alpha \beta_{\dot \alpha}^\alpha + 
\left( \bar k + k \right)  \bar \beta^{\dot \alpha}_\alpha
\beta^\alpha_{\dot \alpha} \\ 
S_{0,1}&=&
 b_\alpha^{\dot \alpha}\bar D_{\dot \alpha} D_\beta
\sigma^{(\beta\alpha)} + \bar\sigma^{(\dot\beta \dot\alpha) \bar
  D_{\dot\beta}} D_\alpha \bar b_{\dot\alpha}^\alpha \,,
\nonumber\\
S'_{*,1}&=&  \bar b^{*\dot \alpha}_\alpha \beta^{\alpha}_{\dot \alpha} +
\bar \beta^{\dot \alpha}_\alpha b^{*\alpha}_{\dot \alpha} \, .
\label{level1}
\end{eqnarray}
The kinetic terms for the superfield $\sigma^\alpha$ and the Lagrange
multiplier $\beta_{\dot \alpha}^{\alpha}$ (and their hermitian
partners) provide a good propagator which can be easily computed.  The
formulation with the Lagrange multiplier is clearly equivalent to the
formulation of \cite{GPZ}. This can be seen by eliminating the
Lagrange multipliers by means of their equations of motion, which are 
\begin{equation}
  \label{eq_lag_mul}
  - \beta^{\alpha}_{\dot \alpha} + \bar D_{\dot \alpha} \sigma^\alpha
  + b^{*\alpha}_{\dot \alpha} = 0,~~~~~~ {\rm for}~~ k  + \bar k =1,   
\end{equation}
and its hermitian conjugate. 

Before discussing the canonical transformations (\ref{Fdiagsigma}), we
note that, due to the trivial BRST variations of the antighosts and
the Lagrange multipliers described in eqs.~(\ref{lag_mul}), it is easy
to eliminate those fields from the BRST cohomology $H^*(\gamma)$ for
zero ghost number. On the other hand, the BRST transformations of the
spinor fields $\sigma^\alpha$ do remove the unphysical degrees of
freedom. At this level, the BRST charge is easily constructed and the
physical subspace does coincide with the classical one.  

Following the suggestions of \cite{GPZ}, we perform the
diagonalization in order to decouple the fields of the $0^{\rm
  th}$-level from the ghost fields in such a way that the further
gauge fixings do not modify the structure of the $0^{\rm th}$-level
field action. The necessary canonical
transformations involve also the Lagrange multipliers:
\begin{eqnarray}
  \label{can_1}
  \sigma^\alpha \longrightarrow  \sigma^\alpha + 
  F(\partial)^{\alpha \dot \tau }_{\tau} b^{*\tau}_{\dot \tau}\, ~~~~~
  \beta^\alpha_{\dot \alpha} \longrightarrow   \beta^\alpha_{\dot \alpha} + 
  G(\partial)^{\alpha \dot \tau }_{\dot \alpha \tau} b^{*\tau}_{\dot
    \tau}\,,   
\end{eqnarray}
and equivalently for the hermitian conjugates. In the above equation,
$F(\partial)^{\dot \tau \alpha}_{\tau}$ and $G(\partial)^{\dot \tau
  \alpha}_{\tau \dot \alpha}$ are integro-differential operators fixed
by cancelling the couplings $S'_{*,1}$ in eq.~(\ref{level1}). Notice
that as a consequence of the canonical transformations (\ref{can_1}),
the antighosts $b^{\alpha}_{\dot\alpha}$ and $\bar b^{\dot
  \alpha}_{\alpha}$ are also modified by
\begin{equation}\label{can_2}
  b^{\dot \alpha}_{\alpha} \longrightarrow b^{\dot
    \alpha}_{\alpha} +  
  \sigma^*_\tau  F(\partial)^{\dot \alpha \tau}_{\alpha}
  + \beta^{*\dot \tau}_\tau  G(\partial)^{\dot  \alpha \,
    \tau}_{\alpha \dot  \tau}\,, 
\end{equation}
where the operators $F(\partial)^{\dot \alpha \tau}_{\alpha},
G(\partial)^{\dot \alpha \, \tau}_{\alpha \dot \tau}$ act on the
right. After inserting the transformed variables in action~(\ref{level1}),
the couplings of the antifields of the
$b^{\alpha}_{\dot \alpha}$ with $0^{\rm th}$-level fields and with
themselves are
\begin{eqnarray} \label{level1_cano1}
S'_{Q,1} + S'_{*,1} \longrightarrow 
&& \bar \sigma^{\dot \alpha} \left( \bar D D  F + D G \right)^{\alpha
  \dot \tau}_{\tau}  
b^{*\tau}_{\dot \tau} + \nonumber \\ 
&& \bar \beta^{\alpha}_{\dot \alpha} 
\left( \bar DF + (k + \bar k) G + 1\right)_{\dot  \alpha \,
  \tau}^{\alpha \dot  \tau} b^{*\tau}_{\dot \tau}  
+ {\rm h.c.} + \\
&& \bar b^{*\dot \alpha}_{\alpha} \left(  \bar F \bar D D F + \bar F D G + 
\bar G \bar D F + (k + \bar k) \bar G G + 
\bar G + G \right)^{\dot  \alpha \, \tau}_{\alpha \dot  \tau}
b^{*\tau}_{\dot \tau}\,. \nonumber  
\end{eqnarray}
In order to avoid a cumbersome notation by making all the spinor
indices explicit, we use the superfield matrix notation
of~\cite{superspace}. The operators $\bar F$ and $\bar G$ are the
hermitian conjugates of $F,G$ plus suitable integrations by parts.  As
a consequence of the canonical transformation, we obtain also the
following terms
\begin{eqnarray}\label{level1_cano2}
  \sigma_{\alpha}^* D_\beta \sigma^{(\alpha \beta)} +  b_\alpha^{\dot \alpha}
\bar D_{\dot \alpha} D_\beta
 \sigma^{(\beta\alpha)} + {\rm h.c.} \longrightarrow 
&& \sigma_{\alpha}^* \left( 1 + F \bar D \right)^{\alpha}_\tau D_\beta 
\sigma^{(\tau \beta)} + \nonumber \\
&& \beta_\alpha^{*\dot \alpha} \left( G \bar D
\right)^\alpha_{\dot \alpha \tau}   
D_\beta \sigma^{(\tau \beta)} + {\rm h.c.}\,.
\end{eqnarray}
Using the solution given in \cite{GPZ}, 
and the relation $1 - \Box^{-1} \Box = 1 - \Box  \Box ^{-1} = K_0$, 
where $K_0$ is the projector on the zero modes of the d'Alembertian
$\Box$, we have  
\begin{eqnarray}
  \label{solu_1}
  F(\partial)^{\dot \tau \alpha}_{\tau} = - {1\over \Box}
  i \partial^{\dot \tau \alpha}  
  D_{\tau}\,,~~~~~~ G(\partial)^{\dot \tau \alpha}_{\tau \dot \alpha} = 
  - \frac{1}{k + \bar k} 
\left( 1 - \frac{i \not\!{\partial}}{\Box} \bar D D \right)^{\dot \tau
  \alpha}_{\tau 
  \dot \alpha}\,. 
\end{eqnarray}
Notice that, treating the integro-differential operators $F(\partial)$
and $G(\partial)$ formally, one misses the contribution of zero modes.
Indeed, only a carefully handling of the definition of the inverse of
$\Box$ introduces the operator $K_0$. We do not need to specify
$K_0$, but we want to underline the fact that it is incorrect to
neglect such contribution.

From eq.~(\ref{level1_cano1}) and setting $k = \bar k = -1/2$, we have 
\begin{eqnarray}
  \label{level1_fin1}
  S'_{Q,1} + S'_{*,1} \longrightarrow 
  \bar \sigma^{\dot \alpha} K_0 D_\tau b^{*\tau}_{\dot \alpha} + {\rm
  h.c.} +   
 \bar b^{*\dot \alpha}_{\alpha} 
\left(  1 + \frac{i\not\!{\partial}}{\Box} \bar D D   (1 - K_0) 
\right)^{\dot  \alpha \, \tau}_{\alpha \dot  \tau} b^{*\tau}_{\dot
  \tau}\, . 
\end{eqnarray}
From eq.~(\ref{level1_cano2}):
\begin{eqnarray}
  \label{level1_fin2}
\sigma_{\alpha}^* D_\beta \sigma^{(\alpha \beta)} +  b_\alpha^{\dot \alpha}
\bar D_{\dot \alpha} D_\beta
 \sigma^{(\beta\alpha)} + {\rm h.c.} \longrightarrow 
&& \sigma_{\alpha}^*  K_0 D_\beta \sigma^{(\alpha \beta)} + \nonumber \\
&& \beta_\alpha^{*\dot \alpha} K_0 \bar D_{\dot \alpha} D_\beta
\sigma^{(\alpha \beta)}  
+ {\rm h.c.}\,.
\end{eqnarray}
We must notice that the integro-differential operators
$F(\partial)$ and $G(\partial)$ cannot cancel all the
possible contributions found in \cite{GPZ}: the contribution
of zero modes is left. For the purpose of the definition of the
propagator, this detail is inessential, but it plays a fundamental role
in the cohomological computation. Indeed, we will show that the
further gauge fixing does not affect the $0^{\rm th}$-level fields (as
expected on the basis of the GPZ technique).

Applying the definition~(\ref{omega}), we have
\begin{eqnarray}
  \label{BRST_non-loc}
  \gamma \, \sigma^\alpha &=& K_0 D_\beta \sigma^{(\alpha \beta)}\,,
  \nonumber \\ 
  \gamma \, b^{\alpha}_{\dot \alpha} &=&  K_0 \bar D_{\dot \alpha}
  \sigma^{\alpha}\,, \\ 
  \gamma \, \beta^{\alpha}_{\dot \alpha} &=& K_0 \bar D_{\dot \alpha}
D_\beta \sigma^{(\alpha \beta)} \,, \nonumber 
 \end{eqnarray}
where the only surviving contributions are the zero modes. This
ensures the nilpotency of the BRST differential $\gamma$ together
with the equations of motion
\begin{eqnarray}
  \label{eq_mo}
  \bar D_{\dot \alpha} D_\beta \sigma^{\beta} +  D_\alpha
  \beta^{\alpha}_{\dot \alpha} = 0\,, ~~~~ 
\bar D_{\dot \alpha} \sigma^{\alpha}  - \beta^{\alpha}_{\dot \alpha} = 0\,. 
\end{eqnarray}
By means of the equations of motion, it is immediate to see that the
degrees of freedom of the Lagrange multiplier depend on
$\sigma^\alpha$. Finally, the non-local BRST transformations are 
enough to show that, among the propagating degrees of freedom, only the
physical ones belong to the cohomology $H^*(\gamma)$. This selects the
correct physical subspace. Notice the non-locality of the BRST
transformations induced by the canonical transformations
(\ref{can_1}) and (\ref{can_2}).

This concludes the analysis of the first level. The other levels can
be studied in the same way. Indeed, one needs the quantized version of
the ghost fields in order to extend the constraints of the gauge
fixing to all orders.
 
Along the lines of the above procedure, one can easily (although
tediously) construct the all-level quantization of the Complex Linear
Superfield. However, already at the second level, cumbersome
expressions hide completely the structure of cohomology and the
computation of the physical spectrum. For that reason we study here a
simpler model, for which the complete (all-levels) action can be easily
computed.


\subsection{A toy model}
\label{toy_model}

The main points of the technique we are using to study infinitely
reducible systems and their cohomology $H(\gamma,H(\delta))$ can be
elucidated on a simple model, which resembles in many respects the
Casalbuoni-Brink-Schwarz
\cite{Brink-Schwarz,Siegel:1983hh,Lindstrom_BS,GH_BS,superparticle}
superparticle, or, more exactly, its fermionic sector. In particular,
the structure of the symmetry and the field content of the two models
are the same, while the details of the interactions between
the various fields account for the non trivial differences. \\
The model is a $0+1$-dimensional theory of a $SO(9,1)$ Majorana-Weyl
spinor $\theta$ (the ten-dimensional Lorentz group is here simply an
internal symmetry, as it is the case for the world-volume theories of
supermembranes), with the classical action
\begin{equation}
  \label{eq:chiral_spinor}
  S_{cl} = - \bar\theta \sigma_- \partial \theta \, .
\end{equation}
It is immediate to see that this model has a gauge symmetry
\begin{displaymath}
  \delta \theta = \sigma_- \kappa_1 \, ,
\end{displaymath}
with an infinite chain of zero modes:
\begin{eqnarray*}
  \delta \kappa_1 & = & \sigma_- \kappa_2 \, , \\
  & \vdots & \\
  \delta \kappa_n & = & \sigma_- \kappa_{n+1} \, , \\
 & \vdots &
\end{eqnarray*}
The gauge symmetry cannot be used to eliminate the unphysical degrees
of freedom of $\theta$ without breaking the $SO(9,1)$ covariance, so
we have to quantize the infinitely reducible theory. The minimal
action can be written at once:
\begin{eqnarray*}
  S_{min} & = & S_{cl} + S_* \, , \\ 
  S_* & = & \bar\theta^* \sigma_- k_1 + \sum_{p \geq 1} \bar
  k_p^* \sigma_- k_{p+1} \, .
\end{eqnarray*}
The gauge-fixing procedure starts with the introduction of non-minimal 
terms 
\begin{displaymath}
  S_{nm,1} = \bar\chi_1^{1*} \pi_1^1
\end{displaymath}
followed by a canonical transformation generated by 
\begin{displaymath}
  \Psi_1 = \bar\chi_1^1 \partial \theta
\end{displaymath}
which yields
\begin{displaymath}
  S_1' = S_{min} + (\bar\chi_1^{1*} - \bar\theta \partial) \pi_1^1 -
  \bar\chi_1^1 \sigma_- \partial k_1 \, . 
\end{displaymath}
We see that the coupling of $\theta$ and $\pi_1^1$ provides a
well-defined propagator for these fields, so that the gauge fixing in
this sector is accomplished, but we have also generated a Lagrangian
for the first ghost $k_1$ which, in turn, needs a gauge fixing. We want
to perform this gauge fixing without affecting the sector already
fixed. This can be done if we are able to
decouple this sector from the ghosts. In particular, since the antifield 
of $\chi_1^1$ will generate other terms which will interact with
$\pi_1^1$, we look for a redefinition of the fields that cancels this
coupling. It turns out that we can not completely cancel it: the
canonical transformation we need for the diagonalization is generated by  
\begin{displaymath}
  \Xi_1 = \bar\theta^* \partial^{-1} \chi_1^{1*} \, .
\end{displaymath}
It corresponds to a redefinition of $\theta$ and $\chi_1^1$. 
The operator $\partial^{-1}$ is the formal inverse of the
derivative. However, since the derivative has a kernel, given by the
constant functions, it can only be inverted on the subspace orthogonal 
to its kernel, so that the following relation holds:
\begin{displaymath}
  \partial \partial^{-1} =  \partial^{-1} \partial = 1 - P_0
\end{displaymath}
where $P_0$ is the projector on the kernel. This has the consequence
that the projected part of the coupling survives; this is enough for
our purposes, since the shifts of the fields due to the gauge-fixings
are always in the form of derivatives, so that they are annihilated by 
the projector. The action after the diagonalization is 
\begin{eqnarray*} \label{S_1''}
  S_1'' & = & - \bar\theta \sigma_- \partial \theta - \bar\theta
  \partial \pi_1^1 - \bar\chi_1^1 \sigma_- \partial k_1 \\
  & + & \bar \theta^* P_0 \sigma_- k_1 + S_{*,2} \\
  & + & \bar\chi_1^{1*} P_0 \pi_1^1 - \bar\chi_1^{1*} (1+P_0) \sigma_- 
  \partial^{-1} \chi_1^{1*} \, .
\end{eqnarray*}
It is worthwhile to analyze the BRST cohomology now.
The equations of motion set all the fields to constant values, or, in
other words, only the projected component under $P_0$ is non-vanishing
on shell. The BRST transformations are
\begin{eqnarray*}
  \gamma \theta & = & P_0 \sigma_- k_1 \, , \\
  \gamma \pi_1^1 & = & 0 \, , \\
  \gamma k_n & = & \sigma_- k_{n+1} \, , \\
  \gamma \chi_1^1 & = & P_0 \pi_1^1 \, .
\end{eqnarray*}
The last transformation shows that $(P_0 \chi_1^1, P_0 \pi_1^1)$ form
a trivial pair. If we missed the contribution of the zero modes, we
would not find at this point the correct physical spectrum, since the
constant part of $\pi_1^1$ would be part of the cohomology.  It should
be noted also that the diagonalization generates non-local terms in
the antifields (the last term in eq. (\ref{S_1''})). These could lead,
after successive gauge-fixings, to non-local terms in the Lagrangian.
This, however, does not happen, as all the potentially dangerous
contributions can be shown to
vanish. \\
The procedure can now be continued, fixing the gauge at any level and
then performing the diagonalization before moving to the next level.
Since our interest does not lie in this particular model, we do not
present the details of the successive diagonalizations, but just point
out that in this case it is not difficult to complete the procedure to
all orders and give the complete action and BRST charge. The result is
as follows:
\begin{eqnarray}
S & = & - \bar \theta \sigma_- \partial \theta - \bar\theta \partial
\pi_1^1 - \sum_{n\geq 1} \sum_{p=0}^m \bar\chi_n^p \partial \pi_{n+1}^{p+1} -
\sum_{n\geq 1} \sum_{p=0}^{[n/2]} \bar\chi_n^p \sigma_- \partial
\chi_n^{p-1} - \sum_{n\geq 1} \bar\chi_{2n}^{2n} \sigma_- \partial
\chi_{2n}^{2n} \nonumber \\
& + & \sum_{n\geq0} \bar k_n^*  \sigma_- P_0 \, (k_{n+1} +
\partial^{-1} \chi_{n+2}^{1*}) + \sum_{n\geq1} \sum_{p=1}^n
\bar\chi_n^{p*} P_0 \pi_n^p \nonumber \\
& - & \sum_{n\geq2} \sum_{p=1}^{[n/2]} \bar\chi_n^{2p*} \sigma_- (1+P_0)
\partial^{-1} \chi_n^{2p-1*} 
- \sum_{n\geq0} \bar\chi_{2n+1}^{2n+1*} \sigma_- (1+P_0)
\partial^{-1} \chi_{2n+1}^{2n+1*}    \, .
\end{eqnarray}


\section{Quantization of the gauged Complex Linear Superfield}
\label{gauged}

The quantization of the linear superfield coupled to gauge fields
has been discussed in refs. \cite{GPZ,PZ}. The methods of these
papers are reviewed in the next section; here, we first review the
idea of P.~Townsend presented in the lectures \cite{3lectures} and we
show how this idea can be implemented in the case of linear superfield
coupled to gauge field.

The coupling of $\Sigma$ to the Yang-Mills superfield $V$ is easily
accomplished by defining covariant derivatives and covariantly linear
superfields with respect to the gauge fields. We use the conventions
in ref.~\cite{superspace}.  Thus we consider $N=1$ superfields $V$,
$\Sigma$ and $\bar\Sigma$ Lie-algebra valued in the adjoint
representation, $V=V^a T_a$, $\Sigma=\Sigma^a T_a$,
$\bar\Sigma=\bar\Sigma^a T_a$ with $[T_a,T_b]=i f^c_{~ab} T_c$ and $tr
T_a T_b= K\delta_{ab}$, and we introduce covariant derivatives, in
vector representation (see \cite{superspace} and
Appendix~\ref{app:susy_gauge}),
\begin{eqnarray*}\label{covderiv}
\nabla_\alpha= e^{-W} D_\alpha e^{W}\,,~~~~~
\bar\nabla_{\dot\alpha} = e^{\bar W} \bar D_{\dot\alpha} e^{-\bar W}\,,~~~~~
\nabla_{\alpha {\dot\alpha}}= -i \{\nabla_\alpha,\bar\nabla_{\dot\alpha}\}\,.
\end{eqnarray*}
According to~ \cite{GPZ,PZ}, in the classical action (suppressing the 
superspace integrals) , the superderivatives are replaced by the 
covariant superderivatives:
\begin{equation}
  \label{gau_1}
  S_{\rm cl}=
{\rm tr} \left(  \bar \sigma^{\dot \alpha} \bar D_{\dot \alpha}
  D_\alpha \sigma^\alpha \right)  
\longrightarrow 
{\rm tr} \left( \bar \sigma^{\dot \alpha}  \bar\nabla_{\dot \alpha}
  \nabla_\alpha \sigma^\alpha \right) 
= 
{\rm tr} \left( \bar \sigma^{\dot \alpha}  e^{\bar W} \bar D_{\dot
    \alpha} e^{-V} D_\alpha e^W \sigma^\alpha \right)\, . 
\end{equation}
The trace is computed over the representation of matter fields
$\sigma$, and $e^V = e^W e^{\bar W}$ defines the real superfield $V$ in
terms of its chiral counterparts. Introducing the redefined matter
fields $\hat\sigma = e^W \, \sigma$ and $\hat {\bar \sigma} = \bar
\sigma \, e^{\bar W}$, we can rewrite the classical action in the form
\begin{equation}
  \label{gau_2}
S_{\rm cl}= {\rm tr} \left(  \hat{\bar \sigma}^{\dot \alpha} \bar
  D_{\dot \alpha} D_\alpha \hat\sigma^\alpha \right)  
+  {\rm tr} \left(  {\hat {\bar \sigma}}^{\dot \alpha} \bar D_{\dot
    \alpha} \left( e^{-V} -1 \right) D_\alpha  
{\hat\sigma}^\alpha \right)\,. 
\end{equation}
The second term is an interaction term starting at order $O(V)$
in the real superfield.  In terms of the redefined fields, the new
minimal action is
\begin{eqnarray}\label{gau_3}
S_{min}&=&S_{\rm cl}+ \sum^{\infty}_{i=1}\hat\sigma_{A_i}^* D_\beta
\hat\sigma^{(\beta A_i)} +   
\sum^{\infty}_{i= 1} \hat\sigma^{(\dot \beta \dot A_i)}  \bar D_{\dot
  \beta} \hat\sigma_{\dot A_i}^*  
+ S_{f,L}+S_{f,R}\nonumber\\
S_{f,L}&=&\hat\sigma^*_{\alpha_1\alpha_2}C^{\alpha_1\alpha_2}\hat\lambda+
\hat\sigma^*_{A_2\alpha_3}
C^{\alpha_2\alpha_3}\hat\lambda^{\alpha_1}\nonumber\\ 
&& +\hat\sigma^*_{A_3\alpha_4} C^{\alpha_3\alpha_4}\hat\lambda^{A_2}
+\hat\varsigma^* C_{\alpha_1\alpha_2}\hat\lambda^{\alpha_1\alpha_2}
+ \ldots \ ,  
\end{eqnarray}
where the zero modes of the gauge transformations $\sigma^{(\alpha_1
  \alpha_2)}, \sigma^{(\alpha_1 \alpha_2 \alpha_3)}, \dots$, are
replaced by the redefined ghost fields $\hat\sigma^{(\alpha_1
  \alpha_2)} ,\hat\sigma^{(\alpha_1 \alpha_2 \alpha_3)}, \dots$.
Notice that, due to the redefinition of fields, the structure of zero
modes of gauge transformations is not changed and the procedure
outlined in the previous sections can be applied without any
modifications.  The only remaining problem is to show that, during the
quantization procedure --which entails the introduction of non-minimal
terms and canonical transformations-- the gauge field $V$ does not
interact with the redefined ghost fields.

After the gauge fixing of the zero level, we noticed that a
redefinition of fields and antifields is useful to cancel the
couplings among the zero-level fields and those of the subsequent
levels.  Applying the canonical transformations in eqs.~(\ref{can_1})
and ~(\ref{solu_1}) to the interaction term in eq.~(\ref{gau_2}), we
find
\begin{eqnarray}
  \label{gau_4}
&&{\rm tr} \left(  {\hat{\bar\sigma}}^{\,\dot \alpha} \bar D_{\dot
    \alpha} \left( e^{-V} -1 \right) D_\alpha  
{\hat\sigma}^\alpha \right) \longrightarrow  \nonumber \\
&&{\rm tr} \left\{ \left(  {\hat{\bar \sigma}}^{\dot \alpha} \bar
    D_{\dot \alpha} \left( e^{-V} -1 \right) D_\alpha  
{\hat\sigma}^\alpha \right) + 
{\hat{\bar b}}^{*\,\dot\alpha}_\alpha  \, {i \partial^{\alpha \dot\tau} \over
    \Box}  \bar D_{\dot\alpha} \bar D_{\dot \tau}
    \left( e^{-V} -1 \right)  D_\beta {\hat\sigma}^\beta +  
\right. \\ 
&& \left. - \hat{\bar \sigma}^{\dot \beta} \bar D_{\dot \beta}
  \, \left( e^{-V} -1 \right)  {i \partial^{\tau \dot\alpha} \over \Box}
   D_{\tau} D_{\alpha}   {{\hat b}}^{*,\alpha}_{\dot\alpha} +
{\hat{\bar b}}^{*\,\dot\beta}_\beta \left[ {\partial^{\beta \dot\rho}
    \over \Box} 
  \bar D_{\dot\beta} \bar D_{\dot\rho} 
  \left( e^{-V} -1 \right)  
{\partial^{\rho \dot\alpha} \over \Box}  D_\rho D_\alpha
\right]  {{\hat
  b}}^{*,\alpha}_{\dot\alpha} 
\right\}\,,
\nonumber
\end{eqnarray}
where the interaction with the gauge field and the antifield $\hat
b^*$ emerges. This is harmless, if there are no additional gauge
symmetries to be fixed. In fact, we have to fix the first level of ghost
by means of the gauge fixing fermion
\begin{eqnarray}
  \label{Psi2}
  \Psi_2 &=&\left( {\hat{\bar d}}^{~({\dot\alpha}\dot\beta)}_\alpha \bar
    D_{\dot\beta}
    {\hat b}^\alpha_{\dot\alpha} +  
    {\hat{\bar \nu}}\, i \, \partial^{\dot\alpha}_\alpha\, {\hat
      b}^\alpha_{\dot\alpha} +  
  {\hat{\bar b}}_{\alpha\beta}^{\dot\alpha} \bar D_{\dot\alpha} \,
  \hat\sigma^{(\alpha\beta)} +  
  {1\over 2}{\hat{\bar \mu}} \, C_{[\alpha\beta]} \,
  \hat\sigma^{[\alpha\beta]} + 
  {\rm h.c.}\right) \nonumber \\ 
  &+& \left( {\hat{\bar d}}^{\,\,(\dot\alpha \dot\beta)}_\alpha\, {\hat
      b}^\alpha_{(\dot\alpha \dot\beta)}  
    + {\hat{\bar \nu}}\, \hat m +  {\hat{\bar
        b}}_{\alpha\beta}^{\dot\alpha} \, \hat 
    \delta^{(\alpha \beta)}_{\dot\alpha} + {\hat{\bar \mu}} \, \hat n
    + {\rm h.c.} 
  \right) \,, 
\end{eqnarray}
This shifts the field $\hat b^*$ with $ \hat b^{*,
  \alpha}_{\dot\alpha} \longrightarrow \hat b^{*, \alpha}_{\dot\alpha}
+ D_\beta \hat d^{(\alpha\beta)}_{\dot\alpha} - i \,
\partial_{\dot\alpha}^\alpha \hat \nu$ and, as a consequence of the
$D$-algebra (cf. \cite{superspace,GPZ}), it generates the new
interaction terms
\begin{eqnarray}
  \label{gau_5}
\left\{ {\hat{\bar \nu}} \left[ (1 - K_0) \bar D^2  \left( e^{-V} -1
    \right) \right] D_\beta {\hat\sigma}^\beta +{\rm h.c.} +  
{\hat{\bar \nu}} \left[ (1-K_0) \bar D^2  \left( e^{-V} -1 \right) D^2
\right]  {{\hat \nu}} 
\right\}\,. 
\end{eqnarray}
These new terms introduce couplings among the gauge field $V$, the
$0^{\rm th}$-level fields and the $2^{\rm th}$-level ghost fields $\hat
\nu$. This can couple the ``classical'' fields with the complete
tower of ghost fields, and thus it would destroy also the quantization
procedure discussed in \cite{GPZ}.

Pursuing the quantization procedure along the lines of \cite{GPZ}, one
has to diagonalize the $1^{\rm st}$ ghost fields (see the table
(\ref{tab1})) in order to eliminate the coupling among antifields and
fields of different levels. It is easy, even though quite long, to show
that, as a consequence of this diagonalization, the field $\nu$
undergoes the following redefinition
\begin{equation}
  \label{gau_6}
 \hat \nu \longrightarrow \hat\nu  - i {1 \over \Box }
 \partial^{\dot\alpha}_\alpha\, \hat b^{*,\alpha}_{\dot\alpha} +  
i {1 \over \Box } D_\beta  \, \partial^{\dot\alpha}_\alpha\,
\hat\beta^{*,(\alpha\beta)}_{\dot\alpha}\,, 
\end{equation}
and its hermitian conjugate transforms accordingly.
The first term of the redefinition of $\nu$ provides the contribution
which, summed to corresponding terms of Eq.~(\ref{gau_4}), cancels the
coupling among $b^*$ and the gauge fields up to zero modes. The second
term does not introduce any further coupling as a consequence
of $D$-algebra. Notice that this already truncates the coupling
between the $0^{\rm th}$-level fields and the ghost tower.

Quantizing the $3^{\rm rd}$ level, one has to define an invertible
kinetic term for the $\hat \nu$ field.
Following \cite{GPZ}, one has to introduce proper Nielsen-Kallosh
ghost fields $\hat \rho_\alpha$ and their antifields $\hat
\rho^*_{\dot\alpha}$. The canonical transformation that fixes the
gauge of $\nu$ and diagonalizes the fields (eliminating the
coupling between $\hat \nu$ and the antifields $\hat
\rho^*_{\dot\alpha}$) amounts to a further redefinition of the $\hat
\nu$ fields of the form
\begin{equation}
  \label{gau_7}
\hat\nu \longrightarrow \hat\nu  + D_\rho \left( H^{\rho \dot\beta} \hat
  \rho^*_{\dot\beta} \right)\,,  
\end{equation}
where $H^{\rho\dot\beta}$ is an integral-differential operator whose
expression is not relevant for our purposes. Again, this shift does
not introduce any new coupling with the gauge fields. From this level
on, no further redefinition or diagonalization can produce interaction
terms with the gauge fields and the quantization procedure of
\cite{GPZ} can be safely applied.

To conclude, we have proved that the quantization procedure of the complex
linear superfield coupled to a gauge field can be performed along the lines
discussed in the previous sections. Here, we stress the fact that, in
order to respect the structure of levels and to avoid coupling among
different types of ghost fields, an initial redefinition of the gauge
field should be performed. The redefined fields show an infinitely 
reducible gauge symmetry of the same form as the non-gauged model. This
resembles the main idea presented by P. Townsend in \cite{3lectures}.


\section{Conclusions}
\label{conclusion}

In the present paper, we reviewed the quantization procedure for
theories with infinitely many ghost fields. In particular, we
considered the complex linear superfield and a simple chiral (toy)
model as examples. The quantization procedure is based on the
conventional Batalin-Vilkovisky formalism, but suitable non-local
canonical transformation are used in order to decouple fields of
different ghost level. The computation of the BRST cohomology
is performed and it is shown that the technique provides the correct
physical degrees of freedom. 
In addition, we considered the complex linear superfield model 
coupled to gauge fields, and we showed that also this model 
can be quantized along the same lines of the ungauged one. 


\medskip
\section*{Acknowledgements.}
\noindent
Research supported in part by NSF grants no. PHY-9722083 and
PHY-0070787.

\newpage


\appendix

\section{Conventions}
 \label{app:conventions}

The superspace conventions we use are the same as in \cite{GPZ}; the
spinor indices are raised with the charge conjugation matrix
$C^{\alpha\beta}$. Some useful relations are: 
\begin{eqnarray}
 D_\alpha\bar D_{\dot \alpha}+
 \bar D_{\dot \alpha}  D_\alpha &=& i\partial_{\alpha\dot \alpha}
 \nonumber\\ 
 D^2&=& {1 \over 2} D^\alpha D_\alpha\nonumber\\
 \Box&=&{1 \over 2} \partial^{\alpha\dot \alpha}\partial_{\alpha\dot \alpha}=
 D^2\bar D^2+\bar D^2 D^2-D^\alpha\bar  D^2D_\alpha\nonumber\\
 D_\alpha D_\beta&=&C_{\beta\alpha}D^2    \label{conv}
\end{eqnarray}

The table (\ref{tab1}) represents the structure of the classical and
ghost fields of the model. It is understood that each table should be
doubled to account for the complex conjugates of all the fields. 
Fields which occur together in the gauge fermion are connected by a
diagonal arrow. 

All the fields and antifields can be assigned an helicity. In the
extended action only fields of different helicity are multiplied
together: any term in the action or in the gauge fermion is of the
form $L\ O\ R$, where $O$ is a possible superspace operator or
spacetime derivative. All fields with upper dotted indices and lower
undotted indices are right-handed, and vice versa for the left-handed
ones. Fields and antifields have opposite helicities, and so do fields
and their complex conjugates.

A delicate point is how one treats the commutation or
anticommutation of fields and the superspace coordinate $\theta^\alpha$
and $D_\alpha$ which are fermionic. The linear superfield $\Sigma$
is bosonic, and, therefore, $\sigma^\alpha$ is fermionic. Since
antifields have opposite statistics to
fields, $\sigma^*_\alpha$ is bosonic. Then, 
all $\sigma$-fields are
fermionic. The gauge fermion is always fermionic.
It follows that $b^\alpha_{\dot \alpha}$ is fermionic and therefore
$\sigma^{[\alpha\beta]}$ is also fermionic, and its ghost $\lambda$ is
bosonic. 

\tabcolsep 1pt
\begin{table}[htb]\caption{Ghost and Antighosts up to third level.}
\label{tab1}\begin{center}\begin{tabular}{ccccccccccccccccc}
F~ &   &   &   &   &   &   & &\ghe{\sigma^{\alpha}}0{-1}&   &   &   &
&   &   & 
 &  \\
 &  &   &   &    &   &   &$\swarrow$ & &    &   &   &   &   &   &   &  \\
F~ &  &      &   &   &   & \ghe{b_{\alpha}^{\dot \alpha}}{-1}0  &   &   &
&\ghe{\sigma^{\alpha_1\alpha_2}}1{-2}&   &
 &   &   &&   \\
 &  &      &   &   & $\swarrow$   &   &   &   & $\swarrow$   &   &   &   &
 &   &  &    \\
F~ &  &   &   &    \ghe{d^{A_2}_{\dot \alpha}}0{-1}  &
\lre     & \ghe{b_{A_2}^{\dot \alpha}}{-2}1  &
 &   &   &\ghe{\sigma^{A_2\alpha_3}}2{-3}     &   &   &\\
B~ &  &   &   &    $\nu$  &
\lre     &$ \mu $ &
 &   &   &$\lambda $ &   &   &\\
 &  &   &   $\swarrow$&   &   &   & $\swarrow$   &   &   &   & $\swarrow$
 &   &      & &   & \\
F~ &  &
\ghe{e_{A_2}^{\dot A_2}}{-1}0 &
&   &   & \ghe{d^{A_3}_{\dot \alpha},d}1{-2}
&  \lre
 & \ghe{b_{A_3}^{\dot \alpha},b}{-3}2  &
 &   &   & \ghe{\sigma^{A_3\alpha_4},\varsigma }3{-4}  &  & \\
B~ &  &
$ \rho^{\dot \alpha}$ &
&   &   & $\nu^{\alpha}$
&  \lre
 & $\mu_{\alpha}$ &
 &   &   & $\lambda^{\alpha}$  &  & 
\end{tabular}\end{center}\end{table}     \tabcolsep 6pt

\tabcolsep 1pt
\begin{table}[htb]\caption{Lagrange multipliers up to third level.}
\label{tab2}\begin{center}\begin{tabular}{ccccccccccccccccc}
F~ &  &   &   &   &   & \ghe{\beta_{\alpha}^{\dot \alpha}}{0}{-1}  &
&   & &  &   &   &   &   & &   \\ 
      &  &   &   &   & $\swarrow$   &   &   &   & &  &   &   &   &   &  &    \\
F~ &  &   &   &    \ghe{\delta^{A_2}_{\dot \alpha}}{1}{-2}  & & & &   & 
\ghe{\beta_{A_2}^{\dot \alpha}}{-1}2     & &   &   & \\
B~ &  &   &   &    $n$  &  & & &   & $ m $    &  &   &   & \\
      &  &   &   $\swarrow$&   &   &   & &    $\swarrow$ &   &   &  &
      &      & &   & \\ 
F~ &  & \ghe{\eta_{A_2}^{\dot A_2}}{0}{-1} & &   &   &
\ghe{\delta^{A_3}_{\dot \alpha},d}{2}{-3} & &&    &   &   & & 
      & \ghe{\beta_{A_3}^{\dot \alpha},b}{-2}3    \\
B~ &  & $ r^{\dot \alpha}$ & &   &   & $n^{\alpha}$ & & & &  &  & & &
$m_{\alpha}$   
\end{tabular}\end{center}\end{table}     \tabcolsep 6pt


\section{BFM and SUSY}
\label{app:susy_gauge}

We recall some features of the BFM in the SUSY context and two
different way to introduce the splitting background-quantum in the
SUSY context. We start from the discussion of the BFM in supersymmetry
provided in \cite{superspace} and we compare with the splitting
discussed in \cite{3lectures}.

The vectorial representation of the gauge superfield is defined by 
the covariant derivatives
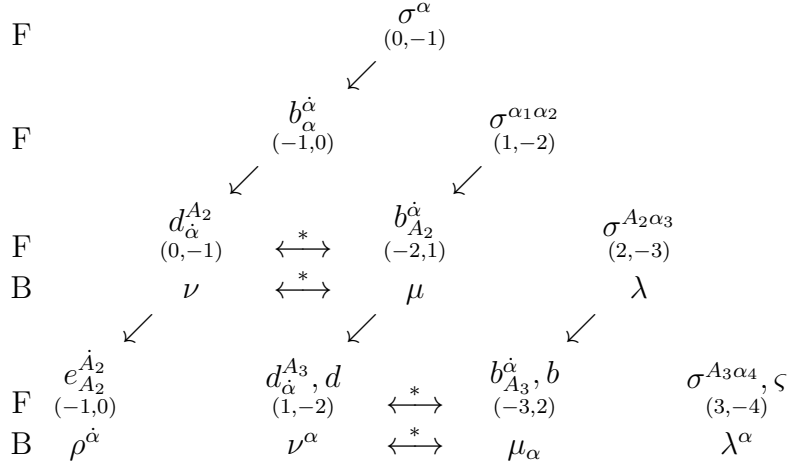
\begin{eqnarray}
  \label{vec_1}
  \nabla^{v}_\alpha = e^{-W} D_\alpha e^W\,,~~~ 
  \bar\nabla^{v}_{\dot\alpha} = e^{\bar W} \bar D_{\bar \alpha}
  e^{-\bar W}\,, ~~~
  \nabla^{v}_\mu = - i \sigma^{\alpha \dot\alpha}_\mu 
  \left\{  \nabla^{v}_\alpha ,   \bar\nabla^v_{\dot\alpha}  \right\}\,.
\end{eqnarray}
These derivatives are covariant with respect to the supergauge
transformations
\begin{eqnarray}
  \label{vec_2}
e^{W'} = e^{i \bar \Lambda} e^W e^{- i K}\,,
~~~  \bar K =K\,,
~~~~ \bar\nabla^v_{\dot \alpha} \Phi^v =0\,.  
\end{eqnarray}
The last equation defines a covariantly chiral superfield $\Phi$.
Another useful representation is the chiral one, which is obtained from
(\ref{vec_1}) by {\it sandwiching} the covariant derivatives of the
vector representation between $e^{-\bar W}$ and $e^{\bar W}$:
\begin{eqnarray}
  \label{ch_1}
&&   \nabla^{c}_\alpha = e^{-\bar W}  \left( e^{-W} D_\alpha e^W \right) 
e^{\bar W} = e^{-V} D_\alpha e^V\,,~~~~  \nonumber 
\bar\nabla^{c}_{\dot\alpha} = \bar D_{\bar \alpha} \,, \\
&&\Phi^{c} = e^{-\bar W} \Phi^v\,, ~~~~ e^V = e^W e^{\bar W}\,.
   \end{eqnarray}
The superfield $V$ transforms under a supergauge transformation
in the following way:
\begin{equation}
  \label{ch_2}
  e^{V'} = e^{W'} e^{\bar W'} = e^{i \bar \Lambda} e^W e^{- i K} e^{ i K}  
e^{\bar W} e^{- i \Lambda} = e^{i \bar \Lambda} e^V e^{- i \Lambda} \,.
\end{equation}

The prepotential $V$ and the complex linear
superfields $\Sigma$'s transform, under gauge transformations, as
\begin{eqnarray}
e^{V'}&=&e^{i\bar{\Lambda}} e^V e^{-i\Lambda}\qquad \qquad 
e^{-V'}=e^{i\Lambda} e^{-V }e^{-i\bar{\Lambda}}
\qquad \qquad \bar D_{\dot\alpha} \Lambda=0~~~~,~~~~ D_\alpha\bar{\Lambda}=0
\nonumber\\
\Sigma'&=&e^{i\Lambda} \Sigma e^{-i\Lambda}\qquad \qquad 
~\bar\Sigma' =e^{i\bar{\Lambda}}\bar\Sigma e^{-i\bar{\Lambda}}
\end{eqnarray}
The action in terms of the new fields is given by
\begin{equation}
S=\frac{1}{\beta^2} \int d^4x~ d^2\theta~ d^2\bar{\theta}~tr\left[-\frac{1}{2}
(e^{-V}D^\alpha e^V)\bar D^2(e^{-V}D_\alpha e^V)-e^{-V} \bar\Sigma e^V \Sigma
\right] \, .
\label{actionprep}
\end{equation}


\end{document}